\documentclass[aps,pra,showpacs,twocolumn,
superscriptaddress,floatfix,footinbib]{revtex4}
    \usepackage{psfig}
    \usepackage{graphicx}
    \usepackage{amssymb}
    \usepackage{amsfonts}
    \usepackage{amsmath}

\newcommand{\veps}{\varepsilon}

\newcommand{\bgar}{\begin{eqnarray}}
\newcommand{\enar}{\end{eqnarray}}
\newcommand{\eq}[1]{(\ref{eq:#1})}

\newcommand{\mc}{\mathcal}

\newcommand{\eqname}[1]{\label{eq:#1}}

\newcommand{\braopket}[3]{ \left\langle #1 \right| #2 \left| #3
\right\rangle }

\newcommand{\expect}[1]{\left\langle #1 \right\rangle}

\newcommand{\x}{ {\bf x}}

\newcommand{\xperp}{ {\bf x}_\perp}
\newcommand{\kperp}{ {\bf k}_\perp}

\newcommand{\al}[1]{^{(#1)}}

\newcommand{\phihd}{\hat\phi^\dagger}
\newcommand{\psihd}{\hat\psi^\dagger}

\newcommand{\psih}{\hat\psi}
\newcommand{\phih}{\hat\phi}
\newcommand{\Eh}{\hat E}
\newcommand{\Ehd}{{\hat E}^\dagger}

\begin{document}
   \title{\bf Imaging of spinor gases}

   \author{Iacopo Carusotto}
   \affiliation{Laboratoire Kastler Brossel, \'Ecole Normale
Sup\'erieure, 24 rue Lhomond, 75231 Paris Cedex 05, France}
   \affiliation{BEC-INFM
    Research and Development Center, I-38050 Povo, Trento, Italy}

   \author{Erich J. Mueller}
   \affiliation{Laboratory of Atomic and Solid State Physics, Cornell
   University, Ithaca, New York 14853}

\begin{abstract}
\end{abstract}

   \date{\today}

\begin{abstract}
We explore the dielectric properties of spinor condensates.
Gases with vector (nematic) order, such as spin-1 condensates with 
ferromagnetic (antiferromagnetic) interactions, display optical activity 
(birefringence).  This optical activity/birefringence is respectively 
seen for light detuned from resonance by a frequency which is not too 
large compared with the fine/hyperfine structure.
Careful tuning of the light frequency can isolate either of these two
effects.
These features can be used to image spin textures in
clouds of spin
1 bosons, and to discern states showing different types of spin
order.
\end{abstract}

   \maketitle

\section{Introduction}

Superfluids display dramatic rotational properties.  For example,
   arrays of quantized
vortices have been observed in rotating gaseous spin-polarized $^{87}$Rb
and $^{23}$Na Bose-Einstein condensates (BECs)~\cite{vortices}.
The physics of rotating systems is even richer in the presence of spin
degrees of freedom: thanks to the interplay of the spin and translational
degrees of freedom, simple vortices are replaced by more complicated
spin textures.
These textures have been studied in many different contexts, from
$^3$He-A~\cite{Helium} to quantum Hall systems~\cite{QuantumHall} and,
more recently, trapped gaseous BECs \cite{SpinTexturesTh,
ErichSpinTexturesTh,spin-1/2Exp,spin-1Exp}.  They are closely related to
three
dimensional skyrmion and hedgehog structures which could be engineered in
clouds of trapped atoms \cite{skyrmion}.

Theoretical studies have
analyzed the structure of spin textures in rotating spinor
BECs for different values of the spin, the trap rotation frequency,
and the atom-atom interactions~\cite{SpinTexturesTh,ErichSpinTexturesTh}.
Experimental studies, beginning with pseudo-spin-$1/2$
systems~\cite{spin-1/2Exp}
and continuing with spin-1 and spin-2 condensates~\cite{spin-1Exp}, have
observed
individual spin textures.
These experiments are complemented by
studies of spin dynamics in non-rotating spin-1 and spin-2
clouds~\cite{spin-2Exp}.

In most of these experiments \cite{exception}, imaging is
performed by first separating  the
different spin components using a magnetic field gradient (the method
pioneered
by Stern and Gerlach \cite{sterngerlach}).  The spin components are then
separately imaged
via conventional optical absorption or phase contrast techniques.
This approach is intrinsically destructive and is unable to
probe coherences between the different
spin components.  Moreover,  images of  the individual
components are very sensitive to global rotations of the
order parameter~\cite{ErichSpinTexturesTh}.

In the present paper, we propose a different imaging scheme which
is predicted to provide non-destructive, {\em in-situ}, images of
the spin textures and which directly addresses the geometrical
features of the BEC order parameter (density, spin, and nematicity
profiles) rather than its components in some specific basis.
The method relies on the
dependence of the dielectric properties of a spinor atomic gas on the
internal state of the atoms.
Focussing our attention on the case of spin-1 atoms, we derive a simple
expression for the local dielectric tensor $\epsilon_{ij}(\x)$ in terms
of the
local density $\rho(\x)$, spin density ${\mathbf S}(\x)$, and
quadrupolar (nematic) order
parameter $N_{ij}(\x)$ of the atomic gas.
Using this expression for $\epsilon_{ij}(\x)$,  we then study the
propagation of polarized light across an atomic cloud: we
show how information about the spatial profile of the spin texture
can be retrieved by analyzing the intensity, phase, and
polarization profile of the transmitted light.
Our technique has analogies in other areas
of condensed matter physics: the polarization of the electrons in a
solid-state sample~\cite{spintronics} as well as the order parameter
of nematic liquid crystals~\cite{liquidcrystals} can be
measured by looking at polarization changes in a transmitted,
diffracted, or reflected light.
As the order parameter of spin-1 atomic systems is a three-component
complex field, it can simultaneously have non-vanishing vector
(spin) and tensor (nematic) components: we discuss possible ways of
isolating
the contribution of each of them.

\section{The light-matter coupling Hamiltonian}

The electric-dipole Hamiltonian describing the
interaction of light with a two-level atom
can be written in a second-quantized form as:
\begin{equation}
    \label{eq:light-matter-1}
    H_{\rm int}=-\int\!d\x\,\sum_{\eta i\alpha}\,D_{\eta
i\alpha}\,\phihd_{\eta}(\x)
\,\Eh_i(\x)\,\psih_{\alpha}(\x).
\end{equation}
The atomic field operators $\phih_{\eta}(\x)$ and $\psih_\alpha(\x)$
destroy an atom at spatial position $\x$ in respectively the $\eta$
sublevel
of the excited state or the $\alpha$ sublevel of the ground state.
$\Eh_i(\x)$ is the $i$-component of the electric field operator at
$\x$ and $D_{\eta i\alpha}=\braopket{e:\eta}{d_i}{g:\alpha}$
gives the matrix element of the $i$-component of the electric
dipole moment  between the sublevels
$\eta$ and $\alpha$ of respectively the excited and ground state.

For light frequencies $\omega$ well detuned from the resonance frequency
   $\omega_0$ of the $g\rightarrow e$ transition,
we can eliminate the excited state and write an effective Hamiltonian
for the
ground state atoms coupled to the
   electromagnetic field:
\begin{equation}
    \label{eq:light-matter-2}
    H_{\rm int}=\int\!d\x\,\sum_{\alpha i j \beta}\,U_{\alpha i j\beta}\,
\psihd_{\alpha}(\x)\,
\Ehd_i(\x)\,\Eh_j(\x)\,\psih_{\beta}(\x),
\end{equation}
where the $U$ tensor is defined as:
\begin{equation}
    \label{eq:U-tensor}
U_{\alpha i j\beta}=\frac{1}{\hbar(\omega-\omega_0)}\,\sum_\eta\,
D^*_{\eta i\alpha}\,D_{\eta j \beta},
\end{equation}
the indices $\alpha$ and $\beta$ run over the ground state
sublevels and $\eta$ runs over the excited state sublevels.

In our case of spin 1 atoms, the matter field is a
three-component vector and can be written in terms of its
(operator-valued) Cartesian components as
${\vec \psi}=(\psih_x,\psih_y,\psih_z)$~\cite{cartesiandef}.
The local one-body density-matrix $m_{ij}(\x)=\psihd_i(\x)\psih_j(\x)$
is the tensor product of the two distinct spin 1 objects ${\vec
    \psi}$ and ${\vec \psi}^\dagger$, so it can be represented in
terms of three
irreducible tensor components
with angular momentum 0, 1 and 2.
Their explicit expressions in Cartesian components are:
\begin{eqnarray}
    \rho(\x)&=&\sum_j m_{jj}(\x)
={\vec \psi}^\dagger(\x)\cdot {\vec \psi}(\x)
    \label{eq:dens} \\
    S_j(\x)&=&-i\,\veps_{jkl}\,m_{kl}(\x)=\left.
-i\, {\vec \psi}^\dagger(\x) \times {\vec \psi}(\x)\right|_j
\label{eq:spin}
    \\
N_{ij}(\x)&=&\frac{1}{2}
[m_{ij}(\x)+m_{ji}(\x)]-\frac{\rho(\x)}{3}\,\delta_{ij}.
    \label{eq:nematicity}
\end{eqnarray}
The totally antisymmetric unit tensor is denoted $\veps_{jkl}$.
Physically, the scalar $\rho$ corresponds to the total density, the
vector
${\mathbf S}$ to the spin density, and the second-rank, symmetric tensor
$N$ to the
{\em nematicity} (or quadrupole moment).
An identical decomposition into the intensity $I$, the spin density
${\mathbf \Sigma}$
and the nematicity ${\mc N}$ can be performed for the electromagnetic
field
one-body density matrix defined as $f_{ij}=\Ehd_i\Eh_j$.

As we assume no external field is present to break rotational
invariance, the Hamiltonian is a scalar (that is spin 0) object.
Scalar objects can be found in the tensor product of two quantities
only if the two quantities have the same spin.
This implies that the Hamiltonian \eq{light-matter-2} can be
decomposed in the form:
\begin{multline}
    \label{eq:light-matter-3}
    H_{\rm int}=\int\!d\x\,\Big[b_0\,I(\x)\,\rho(\x)+b_1\,{\mathbf
        \Sigma(\x)}
\cdot{\mathbf  S}(\x)  \\ +b_2\,\sum_{ij}\,{\mc
N}_{ij}(\x)\,N_{ij}(\x)\Big],
\end{multline}
where the three real coefficients $b_{0,1,2}$ depend on the internal
structure of the atom and on the frequency of the light.  Comparing
(\ref{eq:light-matter-3}) with (\ref{eq:light-matter-2}), one finds
\begin{eqnarray}
b_0&=&\frac{1}{9}\sum_{\alpha\beta} U_{\alpha\beta\beta\alpha}\\
b_1&=&\frac{1}{12}\sum_{\alpha\beta}\left(U_{\alpha\beta\alpha\beta}-
U_{\alpha\alpha\beta\beta}\right)
\\
b_2&=&\frac{1}{6}\sum_{\alpha\beta}\left(U_{\alpha\beta\alpha\beta}+U_{\alpha
\alpha\beta\beta}\right)
-\frac{1}{3}\sum_\alpha U_{\alpha\alpha\alpha\alpha}.
\end{eqnarray}
In the
next section, we give
explicit
expressions for the $U$ tensor and the $b_i$ coefficients for a few 
simple cases.

\section{The dielectric tensor of a spin 1 gas}

The form \eq{light-matter-3} of the light-matter Hamiltonian can be
used for studying the
optical potential induced by light on atoms, as well as the refractive
index
observed by light while crossing the atomic sample.

In the first case, a scalar potential comes from the local intensity
$I(\x)$
of light, a pseudo-magnetic field comes from the electromagnetic spin
${\mathbf   \Sigma}(\x)$, and the ${\mc N}(\x)$ tensor couples to the
nematicity $N(\x)$ of the atoms. A first application of this formalism
has been given in~\cite{Dudarev},
where effects arising from a kind of spin-orbit coupling induced by the
pseudo-magnetic field are analyzed for atoms propagating in
suitably designed optical lattices.

In the present paper, we concentrate on the second
class of phenomena, namely on how the optical properties of the atomic
cloud
depend on the spin state of the atoms; in particular we shall discuss
how the
polarization state of a light beam after crossing the atomic sample can
be
used to obtain information on the spin state of the atoms.

\subsection{The general structure of the dielectric tensor}

Under a mean-field approximation in which the quantum
correlations between the matter and the light field are neglected,
a wave equation for the eigenmodes of the electromagnetic field in a
homogeneous medium can be directly
obtained from the Hamiltonian \eq{light-matter-2}:
\begin{equation}
    \label{eq:el_wave_equation_Schro}
    \big(\hbar\omega\,({\mathbf 1}-\lambda)-\hbar c k P_\perp
    \big)\,{\mathbf E}=0
\end{equation}
where ${\mathbf 1}$ is the identity matrix, the projector $P_\perp$
projects orthogonally to the wavevector
${\mathbf k}$ and the tensor $\lambda$ has been defined as:
\begin{equation}
    \label{eq:lambda}
    \lambda_{jk}=\frac{1}{\epsilon_0}U_{i j k l}\expect{m_{il}},
\end{equation}
where $\epsilon_0$ ($=1/4\pi$ in cgs units) is the permittivity of free 
space.
By comparing this wave equation with the Fresnel equation for a
generic medium of dielectric tensor $\epsilon$:
\begin{equation}
    \label{eq:el_wave_equation_Maxw}
\Big(\frac{\omega^2}{c^2}\epsilon-k^2 P_\perp\Big)\,{\mathbf E}=0,
\end{equation}
one obtains the following explicit expression for the dielectric
tensor of the atomic sample at the frequency $\omega$ in terms of the
coupling tensor $U$ in the Cartesian basis and the one-body density
matrix ${m}$:
\begin{equation}
    \label{eq:epsilon}
    \epsilon_{jk}=\delta_{jk}-\frac{2}{\epsilon_0}\,U_{i j k
l}\,\expect{m_{il}}.
\end{equation}
The same symmetry arguments previously used to parametrize the
Hamiltonian in
the form \eq{light-matter-3} lead to the following expression of the
dielectric tensor \eq{epsilon}:
\begin{equation}
    \label{eq:dielectric_function_2}
    \epsilon_{jk}=\delta_{jk}+c_0 \,\big\langle
\rho\big\rangle\,\delta_{jk}-i
    c_1\,\veps_{jkl}\,\big\langle S_l\big \rangle+
c_2\,\big\langle N_{jk}\big
    \rangle,
\end{equation}
where $c_j=-2b_j/\epsilon_0$.
The scalar term proportional to $c_0$ corresponds to the
isotropic polarizability of
the atoms, while the vector term proportional to $c_1$ describes their
optical
activity around
the axis defined by the atomic spin ${\mathbf S}$. Finally, the tensor
term
proportional to $c_2$ gives birefringence effects, whose principal axis
coincide with the ones of the nematicity ellipsoid defined as:
\begin{equation}
    \label{eq:nemat_ellipsoid}
    (N_{ij}+\frac{\rho}{3}\,\delta_{ij})\,x_i\,x_j=1.
\end{equation}

The matrix elements $D_{ijk}$, and hence the coefficients $c_j$
can be related to the line width $\Gamma$
of the atomic transition \cite{landau}.  For the fundamental $D_{1,2}$
transitions of a typical  alkali atom such as $^{23}$Na or $^{87}$Rb,
$\Gamma\approx 5 \cdot10^7 
\mbox{s}^{-1}$ and 
$|c_j|\approx 10^{-15} \mbox{cm}^{3}\, \Gamma/|\omega-\omega_0|$.
  A typical {\em in-situ} imaging setup \cite{MITDispImag} uses detunings 
in the GHz range (which are large enough that absorption effects can be 
ignored~\cite{Abs_footnote}), 
and atomic densities of order $10^{14} \mbox{cm}^{-3}$, 
resulting in a dielectric tensor which differs  from unity by an amount 
of  magnitude
$||\epsilon-1||\approx 10^{-2}$.  As we shall see in
section~\ref{finestruc}, quantum interference effects 
can drastically reduce the magnitude of any given $c_j$ in some
specific frequency domains.

\subsection{The dielectric tensor for a single transition}

The coefficients $c_i$ in the parametrization
\eq{dielectric_function_2} of the dielectric tensor depend on the
internal structure of the atom.  Since  the ground state
has been assumed to have spin $F_g=1$,
absorbing a photon can bring the atoms into excited states of angular
momentum $F_e={0,1,2}$.
In the present subsection, we shall derive a simple expression of the
$c_i$ coefficients in terms of the strength of the optical transition for
the case where only one excited multiplet is involved.
For notational simplicity we omit the angular brackets denoting
expectation values of matter field operators.

For $F_e=0$, the $D$ tensor in Cartesian components
($j,k=\{x,y,z\}$) has the form
$D_{0 j k}=D_0\,\delta_{j k}$, so the dielectric tensor is:
\begin{equation}
    \label{eq:chi,F=0}
\epsilon_{jk}=\delta_{jk}+A_0\,m_{jk}=\delta_{jk}+
A_0\,\left[\frac{\rho}{3}\,\delta_{jk}
+\frac{i}{2}\veps_{jkl}\,S_l+N_{jk}\right],
\end{equation}
i.e. $c\al{0}_0=A_0/3$, $c\al{0}_1=-A_0/2$, and $c\al{0}_2=A_0$ with
\mbox{$A_0=-2|D_0|^2/\epsilon_0\hbar(\omega-\omega_0)$}.
This result has a simple interpretation: a photon
polarized along the Cartesian axis $j$ interacts only with the component
of the
matter field along the same axis.

For $F_e=1$, the Cartesian ($j,k,l=\{x,y,z\}$) components of the $D$
tensor have the form $D_{j k l}=D_1\,\veps_{j k l}$ and
therefore one has:
\begin{multline}
    \label{eq:chi,F=1}
\epsilon_{jk}=\delta_{jk}+A_1\,\left[\rho\,\delta_{jk}-m_{kj}\right]\\
=\delta_{jk}+
A_1\left[\frac{2}{3}\rho\,\delta_{jk}+\frac{i}{2}\veps_{jkl}\,S_l-
N_{jk}\right],
\end{multline}
i.e. $c_0\al{1}=2 A_1/3$, $c\al{1}_1=-A_1/2$, and $c\al{1}_2=-A_1$ with
\mbox{$A_1=-2|D_1|^2/\epsilon_0\hbar(\omega-\omega_0)$}.
In physical terms, this means that light interacts
only with the matter field polarized along direction orthogonal to the
electric field polarization.

Finally, for $F_e=2$, the dielectric tensor has the form:
\begin{multline}
    \label{eq:chi,F=2}
\epsilon_{jk}=A_2\,\Big[\frac{1}{2}\rho\,\delta_{jk}+
\frac{1}{2} m_{kj}-\frac{1}{3} m_{jk}\Big]
\\
=A_2\left[\frac{5}{9}\rho\,\delta_{jk}-\frac{5i}{12}
\veps_{jkl}\,S_l+\frac{1}{6} N_{jk}\right],
\end{multline}
i.e. $c_0\al{2}=5 A_2/9$, $c\al{2}_1=5 A_2/12$, and
$c\al{2}_2=A_2/6$. The coefficient $A_2$ can be obtained from the
electric dipole moment $D_2$ of the optical transition of the $F_g=1
\rightarrow F_e=2$ transition as
$A_2=-2|D_2|^2/\epsilon_0\hbar(\omega-\omega_0)$.
More specifically, $D_2$ is defined as the electric dipole matrix
element between the $m_F=1$ sublevel of the ground
state and the $m_F=2$ sublevel of the excited state. The matrix
elements of the transitions between the other sublevels are related to
$D_2$ by the appropriate Clebsch-Gordan coefficients.

\subsection{Effect of the fine and hyperfine structures of the atom}
\label{finestruc}

If more than one excited state is involved in the optical process, the
effective coupling $U$ of the ground state atoms to the light is given
by the
sum of the contributions \eq{U-tensor} of each single excited state,
each of
them being weighted by a factor $1/(\omega_i-\omega)$, where $\hbar
\omega_i$ is the
excitation energy and $\omega$ the light frequency. Clearly, the result
will be
dominated by the states which are closest to resonance.
By tuning to frequencies where the contributions from different excited
states cancel, one can take advantage of quantum interference effects
to make at least one of the coefficients $c_{0,1,2}$ vanish.

For example, considerable simplification is found if the detuning is
large compared to either the fine or hyperfine splitting.
Consider the fundamental $nS\rightarrow nP$ transition of an
alkali atom such as $^{23}$Na or $^{87}$Rb: the $S=1/2$ electronic
spin couples to the excited state's $L=1$  electronic orbital angular
momentum to produce two fine components
   of total electronic angular momentum $J=1/2$ or $3/2$. The ground
   state, having instead a vanishing orbital angular momentum $L=0$,
   contains a single fine component of total electronic angular momentum
   $J=1/2$.
Both $^{23}$Na and $^{87}$Rb have a nuclear spin of
   $I=3/2$: coupling between the nuclear and electronic spins therefore
result in
   the ground state splitting into
two hyperfine components with $F=2$ and $F=1$ (in this paper we limit
ourselves to $F=1$).
The $J=1/2,3/2$ fine components of the excited state respectively split
into
two hyperfine components of $F=1,2$, and four hyperfine components of
$F=0,1,2,3$.
For these alkali atoms,
the fine structure separation between the $D_1$
($\mbox{S}_{1/2}\rightarrow \mbox{P}_{1/2}$) and $D_2$
($\mbox{S}_{1/2}\rightarrow \mbox{P}_{3/2}$) lines is of 
the order of THz. 
% Rb D1 -- lambda=794.7 nm, D2 -- lambda=780nm
%        so fine splitting is 7 THz
% Na: 589.0nm and 589.6 nm so fine splitting is 0.5 THz
Hyperfine structure, resulting from the much weaker coupling between the
electron and the nucleus, amounts to a few GHz for the electronic
ground state and to a fraction of GHz for the excited state~\cite{Steck}.
% Rb87 fundamental 6.8Ghz, excited 820Mhz
% Na23 fundamental  1.8Ghz     , excited ??

If the  detuning $\Delta$ of the light is much larger than the hyperfine
splitting $\Delta_{\rm HF}$, the nucleus, whose direct coupling to
radiation is extremely weak,
is not expected to play any role in the dynamics.
In this regime,  the frequency denominators in the contributions
\eq{U-tensor} to the $U$ tensor coming from the different hyperfine
components are approximately equal.
Consequently, $U$ acts as the identity matrix in the space of
nuclear states. It immediately follows that the dielectric tensor
\eq{epsilon} only depends on the electronic part of the atomic
density matrix, $m\al{e}=\textrm{Tr}_n[m]=\sum_{m_I}
\braopket{m_I}{m}{m_I}$, where the index $m_I$ runs over the $2I+1$
possible nuclear spin states.
As the total electronic angular momentum of the ground state is
$J=1/2$, the decomposition analogous to the ones in
\eq{light-matter-3} and \eq{dielectric_function_2} now gives two
components of angular momentum respectively 0 and 1, but no component
of angular momentum 2.
In physical terms, this means that, once the sum over the
hyperfine components of the excited state is performed, one has
$c_2=b_2=0$ and no birefringence effect nor any
mechanical coupling of light to the nematic order parameter $N$ can
be present. The coefficients $c_1$ and $c_0$ will, however, generally be
non-zero~\cite{Clebsch_1/2_3/2}.
For a large but finite detuning as compared with the hyperfine
splitting, the ratio $c_2/c_{1,0}$ scales as $\Delta_{\rm HF}/\Delta$.

If the detuning of the light is also large with respect to the
fine-structure
splitting, then one can neglect the coupling between light and
the electronic spin.  The sum in \eq{U-tensor} then traces out all of
the spin degrees of freedom (both electronic and nuclear), leaving
only the electronic orbital angular momentum.
Consequently, in addition to $c_2=b_2=0$, we have $c_1=b_1=0$, and the
atoms behave as a gas of spherically symmetric scatterers,
with an isotropic dielectric tensor.  In this regime, the
mechanical coupling between light and the atoms
does not depend on their initial state.
This effect is currently exploited in optical traps in
order to obtain a confining potential which traps all spin states in the
same
way~\cite{OpticalTraps}.

\section{The imaging method}\label{imaging}

\begin{figure*}[htbp]
    \begin{center}
    \includegraphics[width=\textwidth]{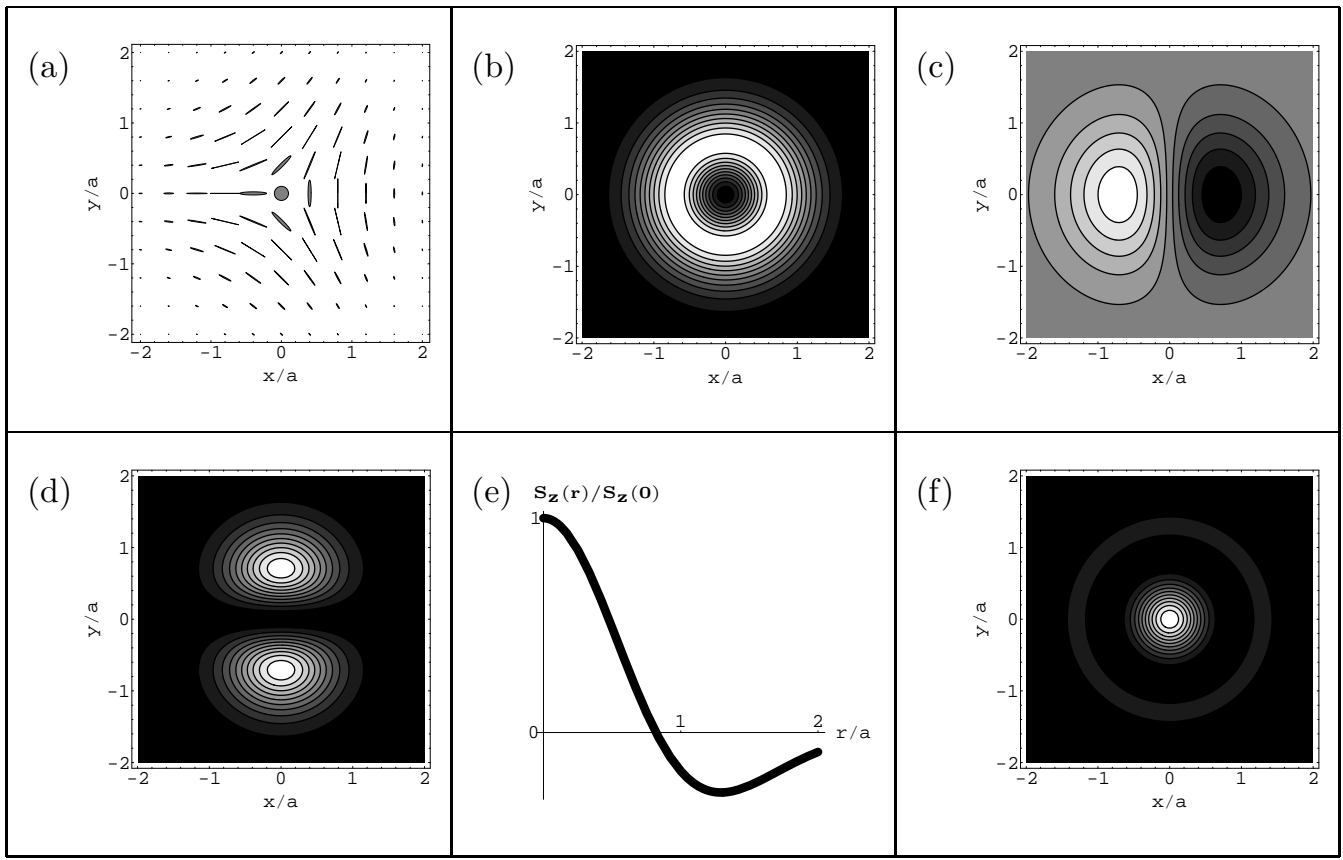}
\caption{
Nematic order of a $\pi$ disclination in a
     rotating spinor BEC with weak antiferromagnetic
      interactions after free expansion for  time $t$.
The lengthscale is given by $a=d_{\rm ho} \sqrt{1+(\omega_{\rm ho} 
t)^2}$, where
$d_{\rm ho}=\sqrt{\hbar/m\omega_{\rm ho}}$,
and $\omega_{\rm ho}$ is the frequency of the harmonic trap in which the
disclination formed \cite{scaling}.
Typically, $a\approx 100\,\mu$m.  If $\omega_{\rm ho}$ is small
enough, then {\em
in-situ} imaging (where $t=0$)
    is possible ($^{87}$Rb in an $\omega_{\rm ho}=10s^{-1}$ trap has 
$d_{\rm
      ho}=9\,\mu$m,
and is amenable to {\em in-situ} imaging).
      Panel (a):  Two dimensional projections of the nematicity ellipsoids
      (\ref{eq:nemat_ellipsoid}) are shown on a regular grid.  Away from
      the center the ellipsoids are degenerate and appear as rods.
      Panel (b,c): Images of the disclination using circularly
      polarized $\sigma_+$ probe light. The intensity of the generated
      $\sigma_-$ component is shown in (b) and a phase-contrast image
after mixing
      with the incident light (with phase $\phi=\pi/2$) is shown in
      (c). Panel (d): Image of the
      disclination using light linearly polarized along $x$ and a crossed
      polarizer, when $c_1=0$. Panel (e): The $\hat z$ component of the
      spin $S_z$, as a function of radial position.  Panel (f):  Image
      of the disclination using linearly polarized light and a crossed
      polarizer when the detuning is much greater than the hyperfine
      splitting so that $c_2=0$.}
      \label{fig:antiferro}
    \end{center}
\end{figure*}

\begin{figure*}[htbp]
    \begin{center}
    \includegraphics[width=\textwidth]{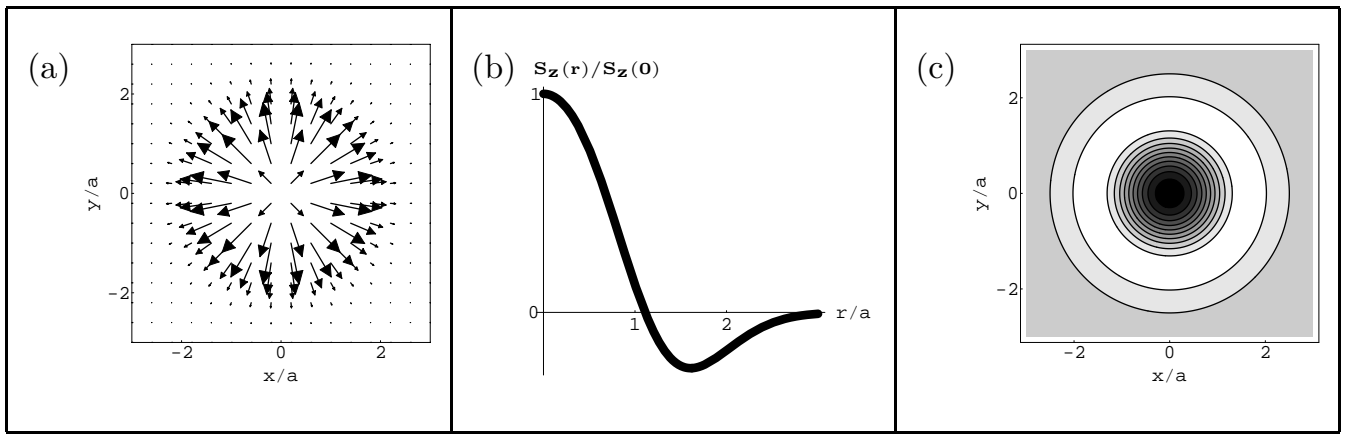}
    \caption{Panel (a,b): geometrical structure of coreless vortex in a
      trapped rotating spinor BEC with weak ferromagnetic
      interactions. Lengthscale $a$ is explained in figure 1.
      In (a), arrows represent the in-plane component of
      the spin, while (b) shows the radial dependance of the $\hat z$
      component.
      Panel (c): image of the coreless vortex by detecting the
      phase difference between the transmitted $\sigma_\pm$ components.}
      \label{fig:ferro}
    \end{center}
\end{figure*}

The simple dependence of the dielectric tensor on
the spin and nematic order parameters \eq{dielectric_function_2} is an
useful
starting point  for optically imaging the order
parameter of a spin 1 atomic sample.
As a simple example, consider a thin, pancake-shaped, Bose-Einstein
condensate which is rotating around the symmetry axis ${\hat z}$.
Depending on the
relative value of the s-wave scattering lengths $a_{0,2}$ in
respectively the singlet and quintuplet channels, the ground state of
the system
in the rotating frame show completely different textures
\cite{ErichSpinTexturesTh}.
As two specific examples, figs.\ref{fig:antiferro}a and \ref{fig:ferro}a
show
the nematicity and spin and patterns for the antiferromagnetic case,
$a_0<a_2$, and the ferromagnetic case, $a_2<a_0$.  In the
former case, the
condensate shows defects such as $\pi$-dislocations in the nematic order
parameter $N$, while the spin ${\mathbf S}$ vanishes outside of the
cores of these defects.
In the latter case, the condensate shows a spin pattern, while the
nematicity ellipsoids are pancake shaped, with their minor axis aligned
with the local spin.

We imagine that the imaging beam propagates along ${\hat z}$, i.e.
parallel to the rotation axis, and
its polarization before interacting with the atoms is given by the
two-component, generally complex, polarization vector ${\mathbf
    p}_{in}=(p_x,p_y)$.
We assume the cloud is optically
thin and that the length scale of the spin pattern is much larger than
the
optical wavelength used for the imaging, so that diffraction effects
during
the propagation through the cloud can be neglected and the atomic density
matrix $m$ can be treated as locally uniform.   These assumptions are 
valid for sufficiently detuned light (for typical atomic densities it 
suffices that $|\omega-\omega_0|>100\,\mbox{MHz}$) and sufficiently weak traps 
($\omega_{\rm ho}\sim 10\,\mbox{Hz}$).  Smaller scale feature can be
imaged by allowing the cloud to expand
ballistically before imaging.

The polarization of the transmitted beam at the transverse position
$\xperp$
after propagation through the cloud is given by the following
expression
in terms of a column integral along the line of sight:
\begin{equation}
    \label{eq:polar_evol}
    {\mathbf p}_{out}(\xperp)=e^{i\omega z/c}\,\left[{\bf
1}+\frac{i\omega}{c}\int\! dz\,\big({n}(\xperp,z)-{\mathbf
1}\big)\,\right]\,{\mathbf p}_{in}.
\end{equation}
The assumption that the cloud is optically thin implies that the
magnitude of the
term involving the
integral is  much smaller than unity.
The matrix giving the local refractive index at the
spatial position $(\xperp,z)$, is defined as the square root
${n}(\xperp,z)=\sqrt{{\epsilon}\al{xy}}$ of the reduced dielectric
tensor ${\epsilon}\al{xy}$ in the $(xy)$ plane, defined as
${\epsilon}\al{xy}={P}_\perp\,{\epsilon}\,{P}_\perp$ where
${P}_\perp$ is the projection operator orthogonal to the ${\hat z}$
axis.
As one can easily see from \eq{dielectric_function_2},
${\epsilon}\al{xy}$
depends on the density $\rho$, on the $S_z$ component of the spin and
on the $N_{xx,yy,xy,yx}$
components of the nematic order parameter only.

If the incident beam is circularly polarized $\sigma_{\pm}$, the density
$\rho$ and the spin component $S_z$ both only give a  phase shift. On the
other hand, the nematicity can mix the two circular polarizations.
We assume that the frequency of the imaging light is chosen so to have
a significant coupling $c_2$ to the nematicity
$N$ (see Appendix). If one illuminates  
the cloud with
pure $\sigma_+$ polarized light of amplitude $\alpha_+\al{0}$, the
$\sigma_-$
component of the field after crossing the cloud has the amplitude:
\begin{equation}
    \label{eq:alpha_-}
    \alpha_-=\alpha_+\al{0}\,\frac{i\omega\,c_2}{4c}
\,\int\!dz\,[N_{xx}-N_{yy}+i(N_{xy}+N_{yx})].
\end{equation}
In geometrical terms, a $\sigma_-$ component is generated as soon as the
(elliptical) cross-section of the nematicity ellipsoid
\eq{nemat_ellipsoid}
along the $z=0$ plane is not circular.
After blocking the original $\sigma_+$ component with a polarizer, the
intensity of the transmitted light $I_-=|\alpha_-|^2$ will be
proportional to
the square of the column integral of the reduced nematicity parameter
\begin{equation}
\|N\|^2=(N_{xx}-N_{yy})^2+\textrm{Re}[2N_{xy}]^2
\end{equation}
along the $xy$ plane. In fig.\ref{fig:antiferro}b we have plotted the
simulated intensity profile for a $\pi$ dislocation in a trapped rotating
antiferromagnetic ($a_2>a_0$) spinor condensate.  In the  $m_z=(1,0,-1)$
basis,
we have taken a condensate wavefunction of the form:
\begin{equation}
\psi(x,y)=\left(
\begin{array}{c}
1 \\ 0 \\ u(x+iy)
\end{array}
\right)\,\exp\Big[-\frac{|x^2+y^2|}{2 d_{\rm ho}^2}\Big]
\eqname{pi_discl}
\end{equation}
with $u=1.22$ and the harmonic oscillator length
$d_{\rm ho}=\sqrt{\hbar/m\omega_{\rm ho}}$ is of the order of
$\mu$m \cite{ErichSpinTexturesTh}.
The corresponding
structure of the nematic order parameter is shown in
fig.\ref{fig:antiferro}a. In this specific case, the magnitude of the
nematicity
$\|N\|$ is symmetric under rotations around the ${\hat z}$ axis.  The
principal axes of $N_{ij}$
rotate by $\pi$ upon circling the disclination center.

More information on the structure of the nematic order parameter
can be extracted by using phase-contrast techniques.
Denoting with $0\leq\theta<\pi$ the angle between the ${\hat x}$ axis
and the
minor axis of the nematicity ellipsoid, it follows from \eq{alpha_-}
that the phase of the generated $\sigma_-$ component makes an angle
$\pi/2+2\theta$ with respect to the incident $\sigma_+$ component.
When this $\sigma_-$ component is mixed with the incident $\sigma_+$
component
(with a mixing phase $\phi$) in a typical phase-contrast
scheme~\cite{PhaseContrast}, the deviation of the detected intensity
from its
background value will be proportional to the quadrature
   $\textrm{Re}[e^{i\phi}{\alpha_-}/{\alpha_+\al{0}}]$.
As an example of such an images, we have plotted in
fig.\ref{fig:antiferro}c a simulated image for the $\phi=\pi/2$ case.  
The
    phase of the generated $\sigma_-$ at points $\x$ and $-\x$ differ by
    $\pi$ as a result of the
    different orientation of the nematic order parameter.
Changing the mixing phase rotates this image by $\phi$.

In experiments on liquid crystals~\cite{liquidcrystals}, one typically
images using linearly polarized light.
A crossed polarizer on the other side of the sample selects out the
orthogonal component of the transmitted light.  If the nematic order
parameter is not parallel or perpendicular to the incident
polarization plane, the polarization plane is rotated, and light will
be transmitted through the crossed polarizer.  The same approach can
be used here, however, the optically active regions where $S_z\neq0$
can also rotate the polarization plane, making it more difficult to
analyze the images.  One can
   isolate the contribution from the
nematic order by working at a frequency at which the coupling of light to
the spin vanishes, {\em i.e.} $c_1=0$.
A simulated image is shown in fig.\ref{fig:antiferro}d for incident
light polarized along $x$: the detected intensity is maximum along the
${\hat y}$ axis where the nematic order parameter makes an angle
$\pm\pi/4$ with
${\hat x}$.

We can isolate the contribution from the atomic spin
${\mathbf S}$  by working at a detuning which is large
compared to the hyperfine structure of the excited state. In this
case, $c_2=0$ and there is no mixing of the $\sigma_\pm$ components.
The difference between the phase shifts of the $\sigma_\pm$ components
results in the rotation of the polarization plane of a linearly
polarized beam by an angle $\theta_{\rm rot}$, which depends on the
column
integral of  $S_z$ as,
\begin{equation}
    \label{eq:theta_rot}
    \theta_{\rm rot}=\frac{\omega\,c_1}{2c}\int\!dz\;S_z.
\end{equation}
The intensity of linearly polarized light passing through  a crossed
polarizer will be proportional to $\sin^2(\theta_{\rm rot})$, which in
the limit
of an optically thin sample is just $\theta_{\rm rot}^2$.  One can
thus get a
direct measurement of the integrated magnitude of the $\hat z$
component of the atomic spins.  In fig.\ref{fig:antiferro}e, $S_z$ is
plotted for the $\pi$-disclination, and in fig.\ref{fig:antiferro}f, a
simulated  image  is shown.

Since $\theta_{\rm rot}$ corresponds to the phase shift between the
$\sigma_\pm$ components, one can also measure it via phase contrast
techniques.
In fig.\ref{fig:ferro}c we show a simulated phase contrast image of a
texture in a ferromagnetic gas.
In the weakly interacting limit \cite{ErichSpinTexturesTh},
this coreless vortex has a wavefunction in the $m_z=(1,0,-1)$
basis  given by:
\begin{equation}
\psi(x,y)=\left(
\begin{array}{c}v(x+iy)^2
\\ u(x+iy)\\ 1
\end{array}
\right)\,\exp\Big[-\frac{|x^2+y^2|}{2 d_{\rm ho}^2}\Big]
\eqname{skyrmion}
\end{equation}
with $u=-1.03$, $v=0.81$, and is illustrated in \ref{fig:ferro}a,b.
Again, $d_{\rm ho}\approx\mu$m is the oscillator length.
In the
external region, the spins points downwards, while they reverse
direction as one approaches the center. Consequently, the image in
fig.\ref{fig:ferro}c shows
a low intensity region at the center, and a higher intensity region
on a ring of radius $1.8 a$ ($=1.8\,d_{\rm ho}$ for {\em in-situ} 
imaging).

\begin{figure*}[htbp]
    \begin{center}
     \includegraphics[width=\textwidth]{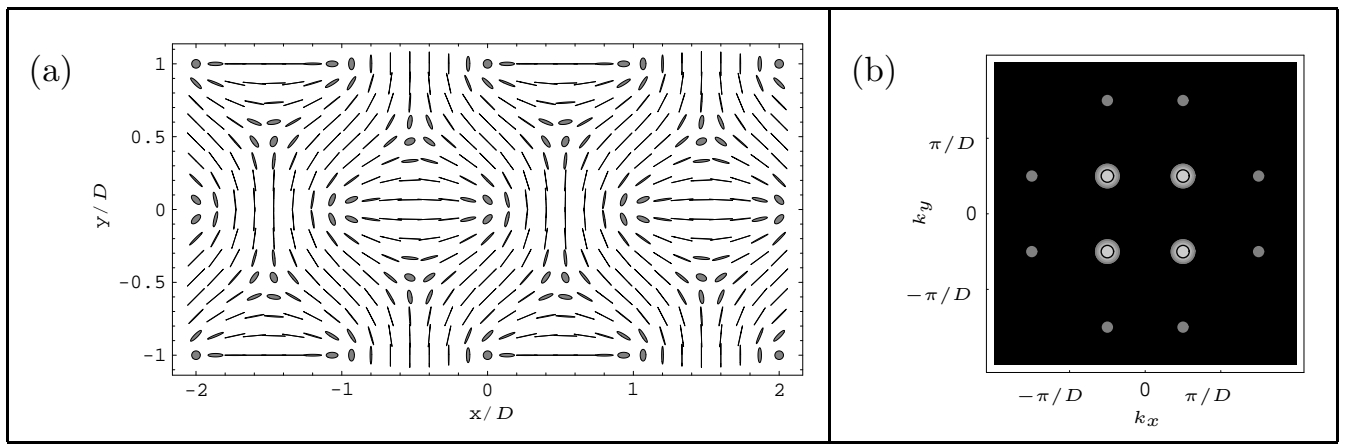}
      \caption{Panel (a): Nematic order in array of $\pi$ disclinations.
      The closest distance between disclinations whose cores have the
      same spin polarization is $D=\sqrt{\hbar/m\Omega}$, where $\Omega$
is the rotation speed
($^{23}$Na rotating at $\Omega=10s^{-1}$ has $D=17\mu$m).
      As in figure~1, this image scales under free expansion.
      Panel (b): Diffraction pattern
      seen in $\sigma_-$ light when $\sigma_+$ light is incident on this
      array of disclinations.  $k_x$ and $k_y$ are the transverse wave
vectors of the diffracted light.}
      \label{fig:Bragg}
    \end{center}
\end{figure*}

If the system shows a repeated pattern of vortices or
$\pi$-dislocations instead of a single one, the imaging may be better
performed in the far-field plane.
The amplitude of the Bragg-diffracted light in the direction
$(\kperp,k_z)$ is proportional to the Fourier transform of the
emerging field after interaction
with the cloud. As an example, we have plotted in fig.\ref{fig:Bragg}b
the Bragg diffraction pattern for the lattice of $\pi$-disclinations
shown in fig.\ref{fig:Bragg}a; the incident light is $\sigma_+$
polarized and $\sigma_-$ diffracted light
is observed. The periodicity of the lattice results in an evenly spaced
series of
isolated diffraction peaks whose strengths and geometrical arrangement
give the detailed geometry of the lattice.

\section{Conclusions}

In summary, we have presented a novel technique for both the {\em
in-situ}
and post-expansion
imaging of spin textures in gaseous spin-1 atomic Bose-Einstein
condensates. The technique is based on the dependence of the
dielectric tensor on the local value of the density and of the spin
and nematic order parameters.
By considering a series of different spin textures, and detection
schemes, we demonstrated how a range of
physical quantities, such as the $z$ component of
the spin, and the magnitude and the orientation of the nematic order
parameter can be imaged.

This polarized imaging technique can also be used to distinguish
between states with and without spin order: both nematic and spin
singlet states have been in fact predicted for antiferromagnetically
interacting spin-1 atoms in an optical lattice~\cite{zhou}.

\acknowledgments

We acknowledge hospitality at the Benasque Center for Science and the
Aspen Center for Physics where most of the present work has been done.
I.C. acknowledges a Marie Curie grant from the EU under contract number
HPMF-CT-2000-00901.
Laboratoire Kastler Brossel is a Unit\'e de
Recherche de l'\'Ecole Normale Sup\'erieure et de l'Universit\'e Paris
6, associ\'ee au CNRS.

\appendix

\section{Quantitative remarks on absorption and the $c_2$ coupling}  

Dispersive imaging techniques, such as this
  one, require that absorption is negligible.  Minimizing absorption
  is also  crucial if one wishes to make {\em in-situ} measurements,
  as the absorption and subsequent spontaneous emission processes heat the
  sample.
  The present imaging scheme of the nematic order parameter requires
  detunings 
  $\Delta=\omega-\omega_0$ not too large than the 
  hyperfine splitting in order to have a significant coupling $c_2$ of
  light to the nematic order parameter. 
 
Since for the $D_1$ line $\Delta_{HF}/\Gamma$ is of the order of
   a few tens~\cite{Steck}, absorption can be avoided while still keeping
  $c_2/c_{0,1}$ of order unity.  These detunings are of the same order as 
those used in the experiments of~\cite{MITDispImag} and the signals are 
therefore readily measured.


\begin{thebibliography}{99}
\bibitem{cartesiandef}  The Cartesian components
    ${\vec \psi}=(\psih_x,\psih_y,\psih_z)$ are related to the more
familiar
    spherical components $(\psih_1,\psih_0,\psih_{-1})$ [where the
    subscript refers to the spin projection in the $\hat z$ direction]
    by
$\psih_x=(\psih_1-\psih_{-1})/\sqrt{2}, \psih_y=i
    (\psih_1+\psih_{-1})/\sqrt{2},\psih_z=\psih_0$.
\bibitem{Abs_footnote} 
Absorption effects are included by changing the energy denominator in
\eq{U-tensor} to
  $\omega-\omega_0-i\Gamma/2$. In the limit of large detuning,
$|\omega-\omega_0|\gg\Gamma$, the $c_i$ coefficients are real 
  and the dielectric tensor is hermitian.
\bibitem{Clebsch_1/2_3/2} After a tedious exercise in Clebsch-Gordon
    coefficients, one finds that for $F=1$ atoms with nuclear spin
    $I=3/2$ $c_1=2 c_0$, while $c_1=-c_0$ for atoms with $I=1/2$.
\end{thebibliography}
\end{document}